\documentclass[aps,twocolumn,superscriptaddress,showkeys,showpacs]{revtex4}
\usepackage{amssymb}
\usepackage{amsfonts}
\usepackage{amsmath}
\usepackage{amsthm}
\usepackage{epsfig}
\usepackage{color}
\usepackage{subfigure}

\newcommand{\eps}{\varepsilon}

\usepackage[normalem]{ulem}

\begin{document}

\title{A Tweezer for Chimeras in Small Networks}

\author{Iryna Omelchenko}
\affiliation{Institut f{\"u}r Theoretische Physik, Technische Universit\"at Berlin, Hardenbergstra\ss{}e 36, 10623 Berlin, Germany}
\author{Oleh E. Omel'chenko}
\affiliation{Weierstrass Institute, Mohrenstra\ss{}e 39, 10117 Berlin, Germany}
\author{Anna Zakharova}
\affiliation{Institut f{\"u}r Theoretische Physik, Technische Universit\"at Berlin, Hardenbergstra\ss{}e 36, 10623 Berlin, Germany}
\author{Matthias Wolfrum}
\affiliation{Weierstrass Institute, Mohrenstra\ss{}e 39, 10117 Berlin, Germany}
\author{Eckehard Sch{\"o}ll}
\email[corresponding author: ]{schoell@physik.tu-berlin.de}
\affiliation{Institut f{\"u}r Theoretische Physik, Technische Universit\"at Berlin, Hardenbergstra\ss{}e 36, 10623 Berlin, Germany}

\date{\today}

\begin{abstract}
We propose a control scheme which can stabilize and fix the position of chimera states in small networks. Chimeras consist of coexisting domains of spatially coherent and incoherent dynamics in systems of nonlocally coupled identical oscillators. Chimera states are generally difficult to observe
in small networks due to their short lifetime and erratic drifting of the spatial position of the incoherent domain. The control scheme, like a tweezer, might be useful in experiments, where usually only small networks can be realized.
\end{abstract}


\maketitle

The study of coupled oscillator systems is a prominent field of research in nonlinear science 
with a wide range of applications in physics, chemistry, biology, and technology.
An intriguing dynamical phenomenon in such systems are {\it chimera states}
exhibiting a hybrid nature of coexisting coherent and incoherent domains~\cite{KUR02a,ABR04,SHI04,LAI09,MOT10,PAN15}.
So far, chimera states have been theoretically investigated in a wide range of large-size networks~\cite{SAK06a,SET08,MAR10,MAR10b,OME11,OME12a,OME12,OME13,HIZ13,PAN13,SET13,SET14,ZAK14,YEL14,SCH14g,XIE14,LAI14,MAI14,MAI15,OME15,PAN15a,LAI15,OLM15,HIZ15,OME15a},
where different kinds of coupling schemes varying from regular nonlocal to completely random topology have been considered.

The experimental verification of chimera states was first demonstrated
in optical~\cite{HAG12} and chemical~\cite{TIN12,NKO13} systems.
Further experiments involved  mechanical~\cite{MAR13}, electronic~\cite{LAR13,LAR15,GAM14}
and electrochemical~\cite{WIC13, SCH14a} oscillator systems
as well as Boolean networks~\cite{ROS14a}.

Deeper analytical insight and bifurcation analysis of chimera states
has been obtained in the framework of phase oscillator systems~\cite{OTT08,OTT09a,OME13a,PAZ14}.
However, most theoretical results refer to the continuum limit only,
which explains the behavior of very large ensembles of coupled oscillators.
In contrast, chimera states in small-size networks have attracted attention only recently~\cite{ASH15,PAN15b,BOE15,HAR16}, 
although in lab experiments usually only small networks can be realized.


There are two principal difficulties preventing the observation of chimera states
in small-size systems of nonlocally coupled oscillators.
First, it is known that chimera states are usually chaotic transients
which eventually collapse to the uniformly synchronized state~\cite{WOL11}.  
Their mean lifetime decreases rapidly with decreasing system size
such that one hardly observes chimeras already for $20-30$ coupled oscillators.
Moreover, a clear distinction between initial conditions that lead to a chimera state,
and those that approach the synchronized state is no longer possible.
Second, the position of the incoherent domain is not stationary
but rather moves erratically along the oscillator array~\cite{OME10a}.
For large systems, this motion has the statistical properties of a Brownian motion
and its diffusion coefficient is inversely proportional to some power of the system size~\cite{OME10a}.
Both effects, finite lifetime and random walk of the chimera position,
are negligible in large size systems.
However, they dominate the dynamics of small-size systems,
making the observation of chimera states very difficult.
To overcome these difficulties some control techniques have been suggested recently.
It has been shown that the chimera lifetime as well as its basin of attraction
can be effectively controlled by a special type of proportional control
relying on the measurement of the global order parameter~\cite{SIE14c}.
On the other hand, Bick and Martens~\cite{BIC15} showed
that the chimera position can be stabilized by a feedback loop
inducing a state-dependent asymmetry of the coupling topology.
However, the latter control scheme relies on the evaluation
of a finite difference derivative for some local mean field.
This operation may become ill-posed for small system sizes like 20--30 oscillators,
therefore one needs to use a refined control in this case.
\begin{figure*}[Ht!]
\includegraphics[height=0.9\linewidth, angle=270]{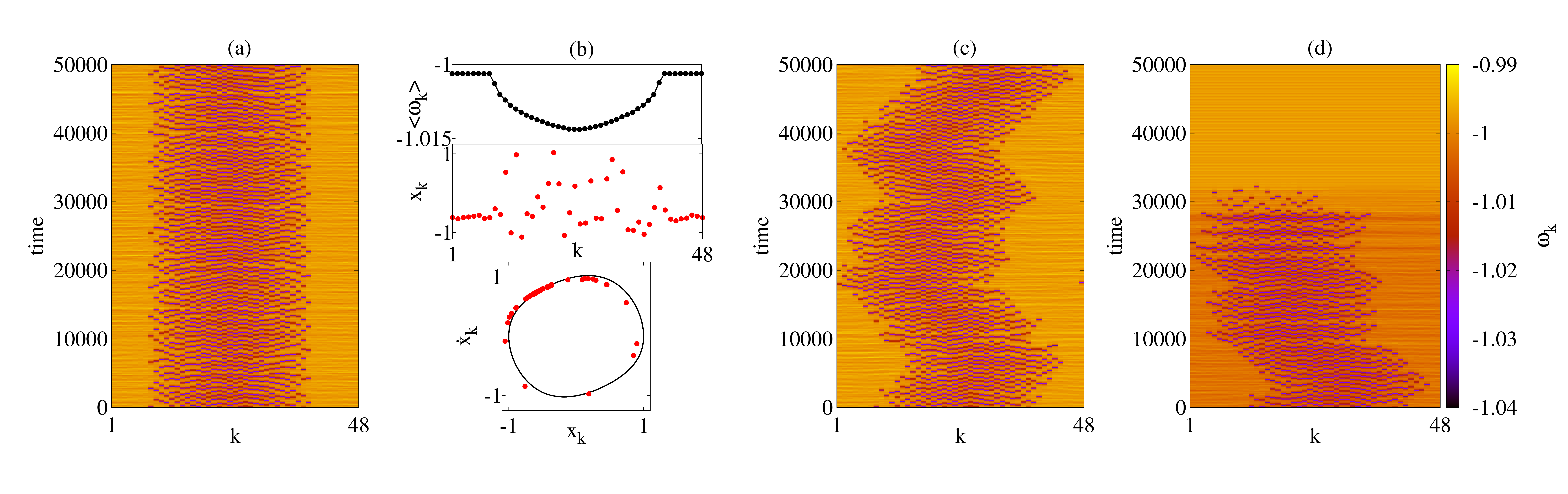}
\caption{(Color online) Mean phase velocities for a system~(\ref{Eq:VdP0})--(\ref{Control})
of $N=48$ oscillators, and $R=16$, $\varepsilon=0.2$, $a=0.02$;
(a)~stable chimera state, $K_{\mathrm{s}} = 0.5$, $K_{\mathrm{a}} = 2$;
(b)~mean phase velocity profile averaged over $\Delta T=50000$ (top panel),
snapshot of variables~$x_k$ (middle panel), and snapshot in the $(x_k,\dot{x}_k)$ phase space at time $t=50000$
(bottom panel, limit cycle of the uncoupled unit shown in black), corresponding to chimera state shown in panel~(a);
(c)~drifting chimera state, $K_{\mathrm{s}} = 0.5$, $K_{\mathrm{a}} = 0$;
(d)~collapse of the uncontrolled chimera state in system~(\ref{Eq:VdP0})--(\ref{Coefficients})
with constant coupling coefficients~$\sigma_\pm$ (see text).}
\label{fig1}
\end{figure*}

In this Letter, we propose an efficient control scheme
which aims to stabilize chimera states in small networks.
Like a tweezer, which helps to hold tiny objects,
our control has two levers: the first one prevents the chimera collapse,
whereas the second one stabilizes its lateral position.
Our control strategy is universal and effective for large as well as for small networks.
Although its justification relies on a phase-reduced model,
the control works also for oscillators exhibiting both phase and amplitude dynamics.

We expect that our tweezer control can also be useful for theoretical studies.
For example, recently Ashwin and Burylko~\cite{ASH15}
introduced the concept of 'weak chimeras', using partial frequency synchronization as the main indicator of such states.
This criterion is slightly different from a 'classical' chimera state,
because the random walk of the incoherent domain results in equal mean phase velocities of all oscillators.
On the other hand, considering chimera states stabilized by the tweezer control,
one can identify them as both 'classical' and 'weak' chimera states.

We consider a system of~$N$ identical nonlocally coupled
Van der Pol oscillators $x_k\in\mathbb{R}$ given by\\[-2mm]
\begin{eqnarray}
\ddot{x}_k &=& (\eps - x_k^2)\dot{x}_k - x_k \nonumber\\[-1mm]
&+& \dfrac{1}{R} \sum\limits_{j=1}^R \left[ a_{-} (x_{k-j}-x_k) + b_{-}(\dot{x}_{k-j} - \dot{x}_k) \right] \nonumber\\[-2mm]
&+& \dfrac{1}{R} \sum\limits_{j=1}^R \left[ a_{+} (x_{k+j}-x_k) + b_{+}(\dot{x}_{k+j} - \dot{x}_k)  \vphantom{\sum} \right].
\label{Eq:VdP0}
\end{eqnarray}
Here, the scalar parameter~$\varepsilon > 0$ determines the internal dynamics of all individual elements.
For small $\varepsilon$ the oscillation of the single element is sinusoidal, while for large $\varepsilon$ it is a
strongly nonlinear relaxation oscillation.
Each element is coupled with~$R$ left and~$R$ right nearest neighbors.
We assume that the oscillators are arranged on a ring
(i.e., periodic boundary conditions)
such that all indices in Eq.~(\ref{Eq:VdP0}) are modulo~$N$.
The coupling constants in position and velocity to the left and to the right
are denoted as~$a_{-}$, $a_{+}$ and~$b_{-}$, $b_{+}$, respectively.
If left and right coupling constants are identical,
i.e., $a_{-} = a_{+}$ and $b_{-} = b_{+}$,
we call the coupling {\it symmetric}, otherwise we call it {\it asymmetric}.
For the sake of simplicity we assume\\[-3mm]
\begin{equation}
a_- = a_+ = a,\qquad
b_- =  a \sigma_-,\qquad
b_+ = a \sigma_+,\\[-2mm]
\label{Coefficients}
\end{equation}
with rescaled coupling parameters~$a$, $\sigma_{-}$ and~$\sigma_{+}$.
Now, combining control approaches from~\cite{SIE14c} and~\cite{BIC15},
we introduce a control scheme for $\sigma_{-}$ and~$\sigma_{+}$, with the aim to stabilize
chimera states of Eq.~(\ref{Eq:VdP0}) not only for large but also for small system sizes.

\begin{figure*}[Ht!]
\includegraphics[height=0.9\linewidth, angle=270]{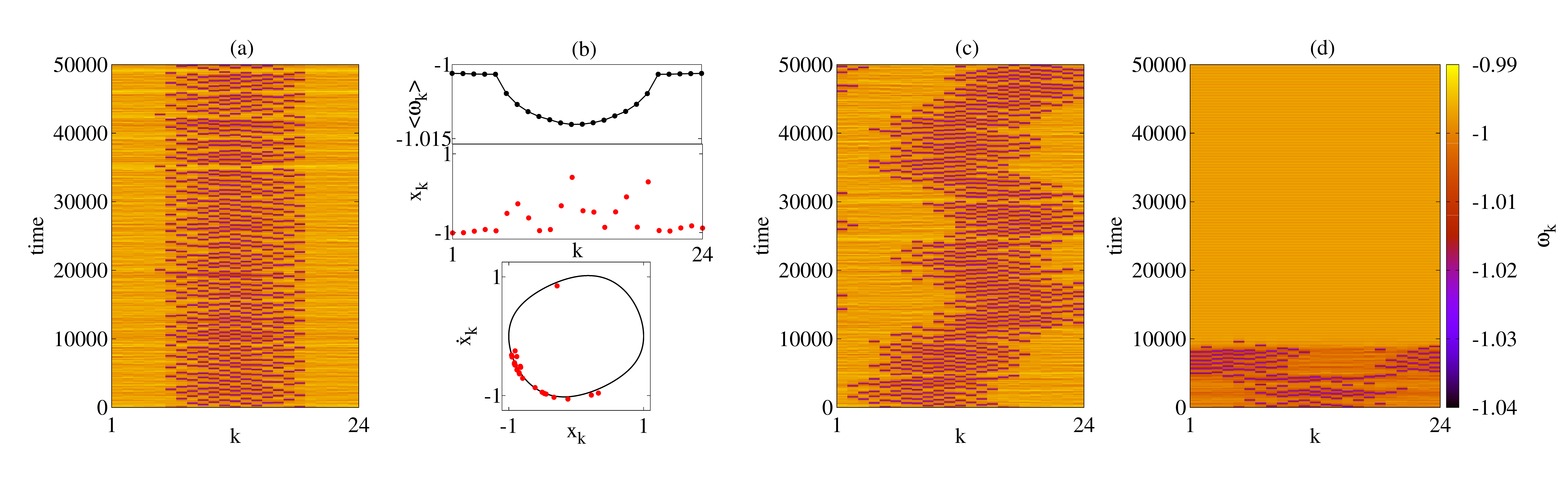}
\caption{(Color online) Same as Fig.~\ref{fig1} for a system~(\ref{Eq:VdP0})--(\ref{Control}) of $N=24$ oscillators and $R=8$.}
\label{fig2}
\end{figure*}

Without loss of generality, we aim to pin the position of the incoherent domain
to the center of the array $1,\dots,N$.
To this end, we define two complex order parameters
\begin{eqnarray}
Z_1(t) &=& \frac{1}{[N/2]} \sum\limits_{k=1}^{[N/2]} e^{i \phi_k(t)}\\
Z_2(t) &=& \frac{1}{[N/2]} \sum\limits_{k=1}^{[N/2]} e^{i \phi_{N-k+1}(t)},
\label{Order_parameter}
\end{eqnarray}
where $\phi_k(t)$ is the geometric phase of the $k$-th oscillator computed from
\begin{equation}
e^{i\phi_k(t)} = \left( x^2_k(t)+ \dot{x}^2_k(t) \right)^{-1/2} \left( x_k(t) + i \dot{x}_k(t) \right).
\end{equation}
Then we define a 'tweezer' feedback control of the form
\begin{equation}
\sigma_\pm = K_{\mathrm{s}} \left( 1-\frac{1}{2}|Z_1 + Z_2|  \right) \pm K_{\mathrm{a}} ( |Z_1| - |Z_2| ),
\label{Control}
\end{equation}
where~$K_{\mathrm{s}}$ and~$K_{\mathrm{a}}$ are gain constants
for the symmetric and asymmetric parts of the control, respectively.
By construction, the quantity $Z = ( Z_1 + Z_2 ) / 2$
coincides with the complex global order parameter,
therefore the feedback terms proportional to~$K_{\mathrm{s}}$
are analogous to the proportional control described in~\cite{SIE14c}.
They suppress the collapse of small-size chimeras,
but do not affect their random drift on the ring.
The latter is the purpose of the terms proportional to~$K_{\mathrm{a}}$.
Indeed, the difference~$|Z_1| - |Z_2|$ measures
a relative shift of the chimera's position
with respect to the center of the array $1,\dots,N$; it is positive if the 
incoherent domain is displaced towards larger indices, and hence $\sigma_{+} > \sigma_{-}$.
On the other hand, a discrepancy between~$\sigma_{-}$ and~$\sigma_{+}$
corresponds to an asymmetry of the coupling
and therefore induces a counterbalancing translational motion of the chimera state.
Thus, for non-zero~$K_{\mathrm{a}}$ a centered configuration
of the chimera state becomes dynamically more preferable.

Fig.~\ref{fig1} illustrates the performance of the suggested control scheme
for the system~(\ref{Eq:VdP0})--(\ref{Control}) of~$N=48$ coupled Van der Pol oscillators.
To visualize the temporal dynamics of the oscillators we plot their mean phase velocities
\begin{equation}
\omega_k(t) = \dfrac{1}{T_0} \int_{0}^{T_0} \dot{\phi}_k(t-t')dt',~~~k=1,\dots,N,
\label{phase_velocity}
\end{equation}
averaged over the time window $T_0 = 50$.
When both the symmetric~$K_{\mathrm{s}}$ and the asymmetric~$K_{\mathrm{a}}$
control gains are switched on (Fig.~\ref{fig1}(a)),
the system develops a stable chimera state
without any spatial motion of the coherent and incoherent domains.
Fig.~\ref{fig1}(b) depicts the mean phase velocity profiles
averaged over the global time window $\Delta T=50000$.
It also shows a snapshot of the chimera for variables~$x_k$
as well as its projection on the phase plane($x_k, \dot{x}_k$).
If we switch off the asymmetric part of the control~$K_{\mathrm{a}} = 0$
and keep a positive symmetric gain~$K_{\mathrm{s}} > 0$,
we find that the chimera state starts to drift (Fig.~\ref{fig1}(c)).
Moreover, if we switch off also the symmetric part of the control
and replace~$\sigma_+$ and~$\sigma_-$ with their effective time-averaged values
$s=\frac{1}{2}(\bar{\sigma}_+ +\bar{\sigma}_-)$ obtained from Fig.~\ref{fig1}(a), 
we find a free chimera state which collapses after some time (Fig.~\ref{fig1}(d)).
Note that the shape of the chimera state is almost unaffected by the control,
which indicates that it is noninvasive on average, cf.~\cite{SIE14c}.

\begin{figure*}[Ht!]
\includegraphics[height=0.9\linewidth, angle=270]{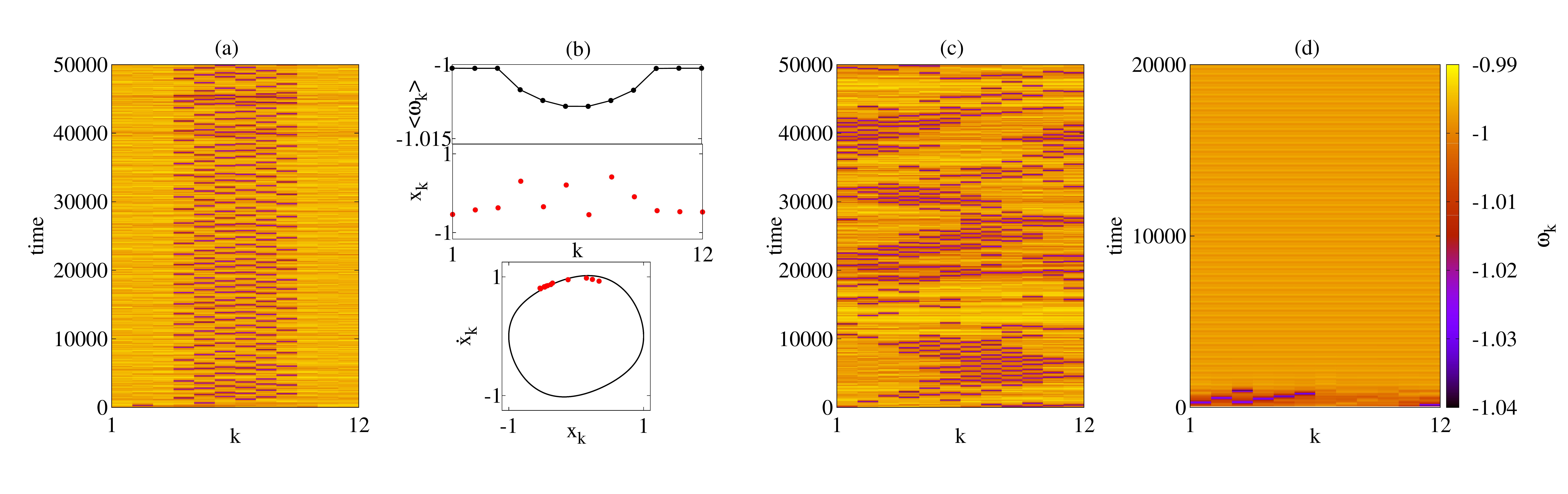}
\caption{(Color online) Same as Fig.~\ref{fig1} for a system~(\ref{Eq:VdP0})--(\ref{Control}) of $N=12$ oscillators and $R=4$.}
\label{fig3}
\end{figure*}

Fig.~\ref{fig2} and Fig.~\ref{fig3} demonstrate that our control scheme
remains effective for smaller networks with~$N=24$ and~$N=12$ oscillators, respectively.
In the controlled system we find chimera states with the same shape
of coherent and incoherent domains (Fig.~\ref{fig2}(b), Fig.~\ref{fig3}(b)).
On the other hand, we observe an increasing difficulty
for chimera states to survive in the uncontrolled system
because of extremely fast wandering (Fig.~\ref{fig2}(c), Fig.~\ref{fig3}(c))
and short lifetimes (Fig.~\ref{fig2}(d), Fig.~\ref{fig3}(d)).

From a phase reduction in the special case of small~$\varepsilon$ and~$a$ (see supplemental material \cite{supp})
one can draw an analogy of the coupling parameters ~$\sigma_\pm$ to the phase lag parameter of the Kuramoto model
$\alpha_\pm = \frac{\pi}{2} - \arctan \sigma_\pm$, and conclude that chimera states should be expected for a range of ~$\sigma_\pm$, where $\alpha_\pm \approx \pi/2$. Note, however, that our control scheme does not rely on the phase reduction of model~(\ref{Eq:VdP0}) to a phase oscillator. Rather, the full model Eq.(\ref{Eq:VdP0}) with phase and amplitude dynamics is used, and the
order parameter is constructed from the geometric phase of the oscillators in the $(x_k,\dot{x}_k)$ phase plane, rather than the dynamic phase of the phase reduced model.

\begin{figure}[Ht!]
\hspace{-0.5cm}\includegraphics[height=\linewidth, angle=270]{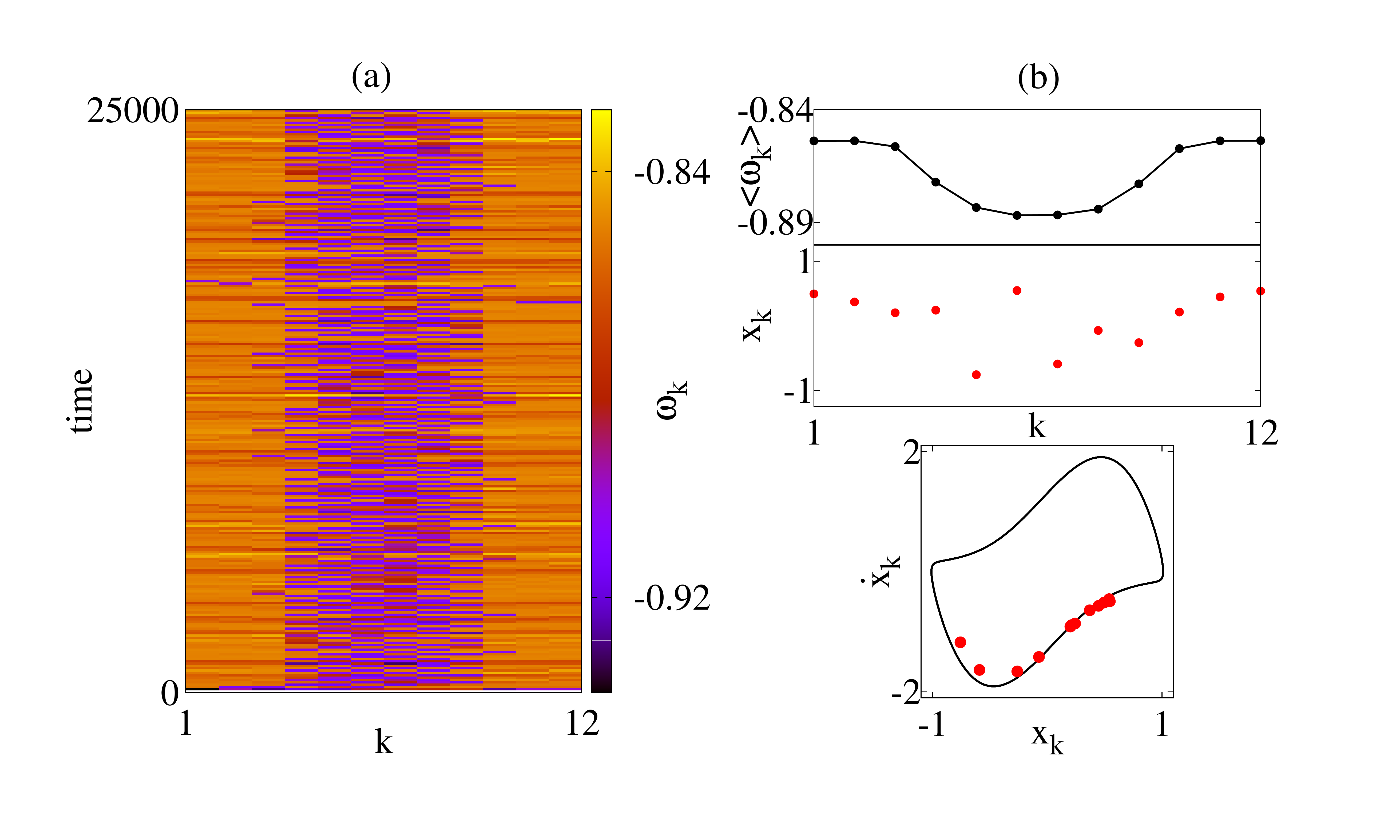}
\caption{(Color online) Same as Fig.~\ref{fig1}(a) for a system~(\ref{Eq:VdP0})--(\ref{Control})
of $N=12$ oscillators, $R=4$, $a=0.2$, $K_s=2$, $K_a=1$, and $\varepsilon = 2$.}
\label{fig_new}
\end{figure}

To substantiate this, we have applied our control scheme to the Van der Pol oscillator in the strongly nonlinear regime of 
non-sinusoidal limit cycles (large $\varepsilon$), see Fig.~\ref{fig_new}. To emphasize the universality of our method, we have
also demonstrated that tweezer control works successfully for small networks of FitzHugh-Nagumo oscillators, i.e.,
coupled slow-fast relaxation oscillators, see Supplemental Material \cite{supp}.
\begin{figure}[Ht!]
\hspace{-0.5cm}\includegraphics[height=\linewidth, angle=270]{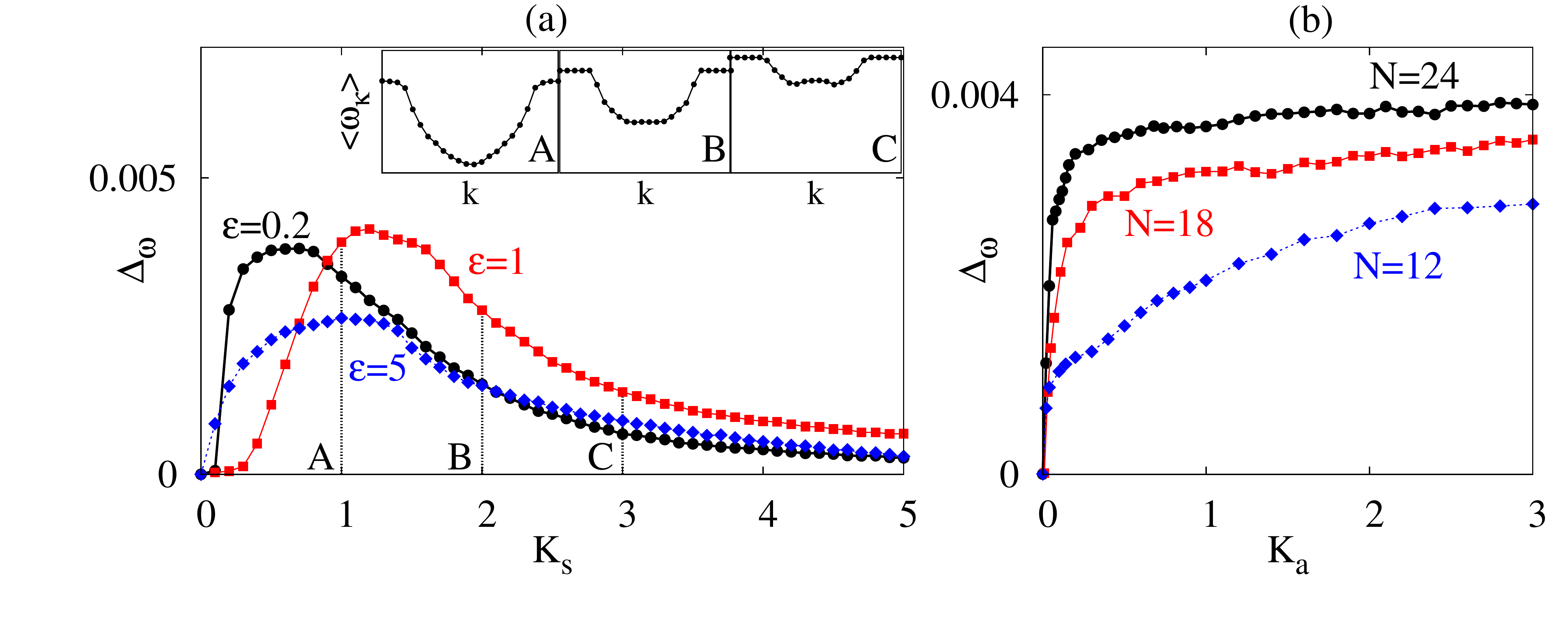}
\caption{(Color online) Standard deviation~$\Delta_\omega$
of the mean phase velocity profiles for $N=24$, $R=8$, $\Delta T=500000$, $a=0.02$.
(a)~Role of the symmetric control strength~$K_{\mathrm{s}}$:
$\varepsilon=0.2$ (black circles), $\varepsilon=1$ (red squares),
$\varepsilon=5$ (blue diamonds), and $K_{\mathrm{a}} = 2$.
Insets show examples of mean phase velocities profiles
for  $\varepsilon=1$ and (A)~$K_{\mathrm{s}}=1$, (B)~$K_{\mathrm{s}}=2$, (C)~$K_{\mathrm{s}}=3$.
(b)~Role of the asymmetry control strength~$K_{\mathrm{a}}$ for different system sizes $N=12, 18, 24$, and
$\varepsilon=0.2$,  $K_{\mathrm{s}} = 0.5$.}
\label{fig5}
\end{figure}

{\it Motivation of control design.}
In order to obtain insight into the mechanism of our control scheme,
we introduce the standard deviation of the mean phase velocity profile 
$\Delta_{\omega}= \sqrt{\dfrac{1}{N}\sum\limits_{k=1}^N (\omega_k - \overline{\omega})^2}$,
where $\overline{\omega}=\dfrac{1}{N}\sum\limits_{k=1}^N \omega_k$.
Larger values of~$\Delta_{\omega}$ correspond to a well pronounced
arc-like mean phase velocity profile, characterizing chimera states.
Fig.~{\ref{fig5}} depicts the influence of the control parameters~$K_s, K_a$ 
on the chimera behavior in the network of $N=24$ oscillators.

First note that, generally, large coupling constants favor complete synchronization. Therefore increasing the symmetric control coefficient $K_s$ has two counteracting effects: On the one hand, increasing $K_s$ increases the global order parameter, which is for even $N$ given by $|Z|=\frac{1}{2}| Z_1 + Z_2 |$, where $|Z|=1$ corresponds to complete synchronization. Hence, on the other hand, the symmetric control term $(1-|Z|)$ in Eq.(\ref{Control}) decreases, i.e., the coupling constants $b_- =  a \sigma_-,b_{+}=  a \sigma_+$ in Eq. (\ref{Eq:VdP0}) decrease. Thus there exists an optimal value of $K_s$ where $K_{\mathrm{s}} \left(1-\frac{ | Z_1 + Z_2 | }{2}\right)$ is optimum, and the chimera state is stabilized; for smaller $K_s$ the control is not efficient, and for larger $K_s$ complete synchronization dominates. To visualize this mechanism better, we have plotted in the inset of Fig.~\ref{fig5}(a) the spatial profiles of the mean phase velocity A, B, C corresponding to three different values of $K_
s$. Roughly speaking, for larger values of $K_s$ we obtain chimera states with larger coherent domains which are closer to the completely coherent state. On the other hand, for smaller values $K_s$ we obtain chimeras with a dominating incoherent domain. Thus, there exists a range of $K_s$ where the arc-shape of the mean phase velocities (see panels (b) in Figs.~\ref{fig1}-\ref{fig_new}) is most pronounced and $\Delta_{\omega}$ is maximum. Such chimeras are most likely to be stabilized in small $N$ systems, therefore this explains the optimal symmetric gain $K_s$. 

Fig.~{\ref{fig5}}(a) shows the effect of the symmetric control~$K_{\mathrm{s}}$
for three values of $\varepsilon = 0.2$, $1$, $5$.
Varying~$K_{\mathrm{s}}$ we stabilize chimera states
with different mean phase velocity profiles as shown for the exemplary case $\varepsilon = 1$ in insets~A, B and~C.
For each $\varepsilon$ there exists a range of parameter~$K_{\mathrm{s}}$,
where the symmetric control is most efficient:
for large~$K_{\mathrm{s}}$ chimera states approach the completely synchronized state ($\Delta_\omega=0$).
In Figs.~\ref{fig1}--\ref{fig3}, we have chosen $K_{\mathrm{s}} = 0.5$ for $\varepsilon=0.2$,
close to the maximum of the black circles.
Increasing $\varepsilon$ from $0.2$ to $1$ (red squares) increases the amplitude of the limit cycle, and hence larger coupling strengths are required, and the maximum of $\Delta_{\omega}$ shifts to larger $K_{\mathrm{s}}$. For very large $\varepsilon$ (blue diamonds), 
$\Delta_{\omega}$ generally decreases.
Fig.~{\ref{fig5}}(b) demonstrates the effect of changing
the asymmetric control gain~$K_{\mathrm{a}}$ for different system sizes.
The standard deviation~$\Delta_{\omega}$ sharply increases for small values of the control strength,
and then stays approximately at the same value indicating the saturation of the position control.
Therefore, in our example $\varepsilon = 0.2$ we choose $K_{\mathrm{a}} = 2$.

The standard deviation of the mean phase velocity profiles increases montonically with system size and saturates at moderate sizes, as shown in the Supplemental Material \cite{supp}. 
Other control scheme is also robust with respect to variation of the nonlinearity parameter~$\varepsilon$ 
and with respect to inhomogeneities of $\varepsilon$, corresponding to an inhomogeneous frequency distribution,
see additional figures in the supplemental material \cite{supp}. 

To conclude, we have proposed an effective control scheme,
which allows us to stabilize chimera states in large and in small-size networks.
Our control is an interplay of two instruments, the symmetric control term
suppresses the chimera collapse, and the asymmetric control
effectively stabilizes the chimera's spatial position.
We have demonstrated the effect of the control scheme
in systems of~$48$, $24$, and~$12$ nonlocally coupled Van der Pol oscillators,
and investigated the role of system parameters and control strengths
for the most efficient stabilization of chimera states.
Our proposed approach can be useful for the experimental realizations of chimera states,
where usually small networks are studied,
and it is very difficult to avoid chimera collapse and spatial drift.

This work was supported by Deutsche Forschungsgemeinschaft
in the framework of Collaborative Research Center SFB~910.


\clearpage

\onecolumngrid

\section*{Supplemental Material on}
\begin{center}
{\large\bf A Tweezer for Chimeras in Small Networks}\\[5mm]
Iryna Omelchenko$^1$, Oleh E. Omel'chenko$^2$, Anna Zakharova$^1$, Matthias Wolfrum$^2$, and Eckehard Sch{\"o}ll$^1$\\[4mm]
$^1${\it Institut f{\"u}r Theoretische Physik, Technische Universit\"at Berlin, Hardenbergstra\ss{}e 36, 10623 Berlin, Germany}\\
$^2${\it Weierstrass Institute, Mohrenstra\ss{}e 39, 10117 Berlin, Germany}\\[8mm]
\end{center}
\twocolumngrid

\subsection{Phase reduction of the main model}
\numberwithin{equation}{subsection}
\setcounter{equation}{0}

Let us denote
$$
x_k(t) = \sqrt{\varepsilon} u_k(t),\qquad \dot{x}_k(t) = \sqrt{\varepsilon} v_k(t),
$$
then the original system of $N$ coupled Van der Pol oscillators Eq.~(1)
can be rewritten as a $2N$-dimensional dynamical system of the form
\begin{eqnarray}
\dot{u}_k &=& v_k,\nonumber\\[2mm]
\dot{v}_k &=& \varepsilon (1-u_k^2)v_k - u_k \nonumber\\[1mm]
& + & \dfrac{a}{R} \sum\limits_{j=1}^R \left[ (u_{k-j} - u_k) + \sigma_{-}(v_{k-j} - v_k) \right] \nonumber\\[1mm]
& + & \dfrac{a}{R} \sum\limits_{j=1}^R \left[ (u_{k+j} - u_k) + \sigma_{+}(v_{k+j} - v_k)  \vphantom{\sum} \right].
\label{System:UV_app}
\end{eqnarray}

We perform a phase reduction in order to determine a parameter set appropriate for observation of chimera states.

Assuming that~$\varepsilon$ and~$a$ are both small,
we can apply the averaging procedure to Eqs.~(\ref{System:UV_app}).
To this end we substitute the ansatz
$$
u_k = r_k \sin(t+\theta_k),\qquad
v_k = r_k \cos(t+\theta_k)
$$
into system~(\ref{System:UV_app}) and average out the fast time~$t$,
assuming that amplitude~$r_k(t)$ and phase~$\theta_k(t)$ are slowly varying functions.
As result we obtain the system
\begin{eqnarray*}
\dot{r}_k &=& \dfrac{\eps}{8} r_k (4-r_k^2) \\[1mm]
&+& \dfrac{a}{2R} \sum\limits_{j=1}^R  \left[ r_{k-j} \sin(\theta_{k-j} - \theta_k) \right. \\
& & \hphantom{\dfrac{a}{2R} \sum\limits_{j=1}^R} \left.  + ~\sigma_{-}(r_{k-j}\cos(\theta_{k-j} -\theta_k) - r_k) \right]\\[1mm]
&+& \dfrac{a}{2R} \sum\limits_{j=1}^R  \left[ r_{k+j} \sin(\theta_{k+j} - \theta_k) \right. \\ 
& & \hphantom{\dfrac{a}{2R} \sum\limits_{j=1}^R} \left.  + ~\sigma_{+}(r_{k+j}\cos(\theta_{k+j} -\theta_k) - r_k) \right], \\
\end{eqnarray*}
\begin{eqnarray*}
r_k \dot{\theta}_k &=& \dfrac{a}{2R} \sum\limits_{j=1}^R \left[ - (r_{k-j}\cos(\theta_{k-j} - \theta_k) - r_k)  \right. \\
& & \hphantom{\dfrac{a}{2R} \sum\limits_{j=1}^R} \left.  - ~\sigma_{-} r_{k-j} \sin(\theta_k - \theta_{k-j})\right] \\[1mm]
&+& \dfrac{a}{2R} \sum\limits_{j=1}^R \left[ - (r_{k+j}\cos(\theta_{k+j} - \theta_k) - r_k)  \right. \\
& & \hphantom{\dfrac{a}{2R} \sum\limits_{j=1}^R} \left.  - ~\sigma_{+} r_{k+j} \sin(\theta_k - \theta_{k+j})\right]
\end{eqnarray*}
which can also be rewritten as follows
\begin{eqnarray}
\dot{r}_k &=& \dfrac{\eps}{8} r_k \left(\left( 4-\dfrac{4a}{\eps}(\sigma_{-} + \sigma_{+}) \right) - r_k^2 \right) \\[2mm]
&+& \dfrac{a}{2R} \sqrt{1+\sigma_{-}^2} \sum\limits_{j=1}^R r_{k-j} \cos (\theta_k - \theta_{k-j} + \alpha_{-}) \nonumber \\[2mm]
&+& \dfrac{a}{2R} \sqrt{1+\sigma_{+}^2} \sum\limits_{j=1}^R r_{k+j} \cos (\theta_k - \theta_{k+j} + \alpha_{+}), \nonumber
\label{Rdot}
\end{eqnarray}
\begin{eqnarray}
\dot{\theta}_k &=& a  - \dfrac{a}{2R}\sqrt{ 1 + \sigma_{-}^2}\sum\limits_{j=1}^R \dfrac{r_{k-j}}{r_k}\sin(\theta_k - \theta_{k-j} + \alpha_{-}) \nonumber\\[2mm]
&\phantom{=}& \phantom{a} - \dfrac{a}{2R}\sqrt{ 1 + \sigma_{+}^2}\sum\limits_{j=1}^R \dfrac{r_{k+j}}{r_k}\sin(\theta_k - \theta_{k+j} + \alpha_{+}),\nonumber
\label{ThetaDot}
\end{eqnarray}
If $0<a \ll\eps$, from Eq.~(A.2)
we find that~$r_k\approx 2$ is a stable fixed point.
Substituting this into the second equation, we obtain a Kuramoto-like system
\begin{eqnarray}
\dot{\theta}_k &=& a  - \dfrac{a}{2R}\sqrt{ 1 + \sigma_{-}^2}\sum\limits_{j=1}^R \sin(\theta_k - \theta_{k-j} + \alpha_{-}) \nonumber\\[-1mm]
&\phantom{=}& \phantom{a} - \dfrac{a}{2R}\sqrt{ 1 + \sigma_{+}^2}\sum\limits_{j=1}^R \sin(\theta_k - \theta_{k+j} + \alpha_{+})
\label{System:Kuramoto}
\end{eqnarray}
where
\begin{equation}
\alpha_\pm = \mathrm{arccot}\:\sigma_\pm = \frac{\pi}{2} - \arctan \sigma_\pm.
\label{Formula:alpha_sigma}
\end{equation}
Note that for~$\sigma_{-} = \sigma_{+}$, equation~(\ref{System:Kuramoto})
is equivalent to the system considered in~\cite{S_OME10a,S_WOL11,S_WOL11a}.
This suggests a range of parameters~$\sigma_\pm$
where chimera states should be expected, i.e., $\alpha_\pm \approx \pi/2$.

\subsection{Role of nonlinearity and system size}
\numberwithin{equation}{subsection}
\setcounter{equation}{0}
\begin{figure*}[Ht!]
\includegraphics[height=0.8\linewidth, angle=270]{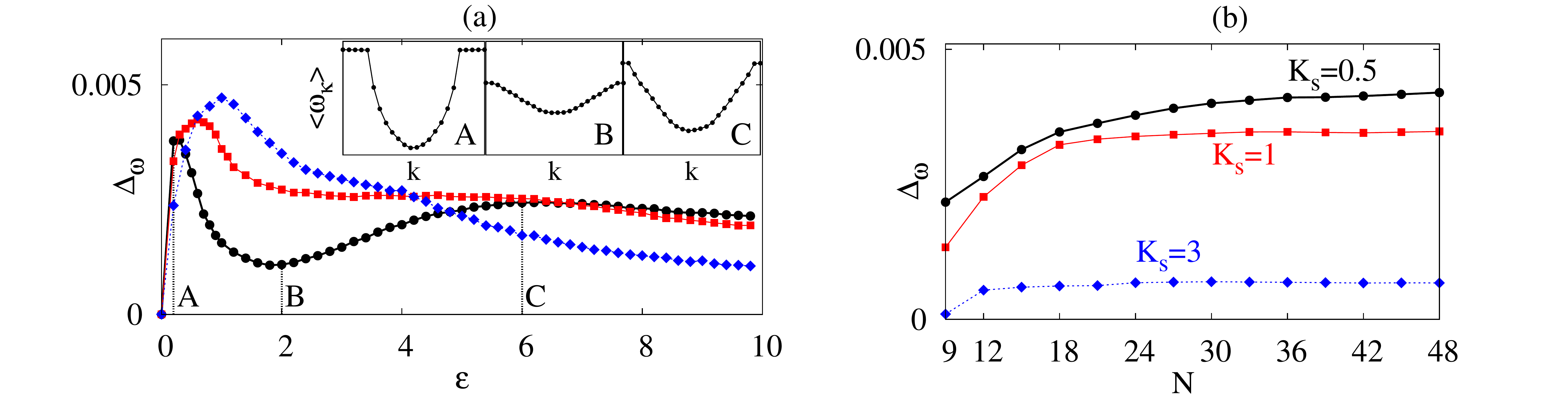}
\caption{(Color online) Standard deviation of the mean phase velocity profiles
for $\Delta T=100000$, $a=0.02$.
(a) Effect of parameter~$\varepsilon$ for~$K_s=0.5$ (black circles), 
$K_s=1$ (red squares), $K_s=1.5$ (blue diamonds), and $K_a=2$, $N=24$, $R=8$.
Insets show examples of mean phase velocities profiles
for $K_s=0.5$ and (A)~$\varepsilon=0.2$, (B)~$\varepsilon=2$, (C)~$\varepsilon=6$;
(b)~Role of the system size~$N$: $K_s=0.5$ (black circles),
$K_s=1$ (red squares), $K_s=3$ (blue diamonds), and $\varepsilon=0.2$, $K_a=2$, $r=R/N=1/3$. }
\label{fig_measures}
\end{figure*}

To analyse the influence of the system parameters on the controlled chimera states,
we use the standard deviation of the mean phase velocity profile 
$\Delta_{\omega}= \sqrt{\dfrac{1}{N}\sum\limits_{k=1}^N (\omega_k - \overline{\omega})^2}$,
where $\overline{\omega}=\dfrac{1}{N}\sum\limits_{k=1}^N \omega_k$.
Larger values of~$\Delta_{\omega}$ correspond to a well pronounced
arc-like mean phase velocity profile, characterizing chimera states.
Fig.~{\ref{fig_measures}} depicts the influence of the nonlinearity parameter~$\varepsilon$
of the individual Van der Pol unit and of system size~$N$ on the mean phase velocity profiles.

Increasing~$\varepsilon$ results in changing the dynamics of the individual elements 
from regular sinusoidal oscillations to relaxation oscillations.
Fig.~{\ref{fig_measures}}(a) shows that for small values of~$\varepsilon$,
the chimera states are well pronounced (inset A, black circles denoting~$K_s=0.5$),
for intermediate values the difference
between maximum and minimum phase velocity is very small (inset B),
and for even larger~$\varepsilon$ it increases again (inset C).
When symmetric control becomes stronger
($K_s=1$ or $1.5$, shown by red squares and blue diamonds, respectively)
again for intermediate values of~$\varepsilon$ the chimera states are more pronounced,
while larger nonlinearity results in a decrease of~$\Delta_{\omega}$, but the maximum is shifted to larger $\varepsilon$, since
for larger $\varepsilon$ the amplitude of the limit cycle grows, and larger $K_s$ matches better with optimum control.

Fig.~{\ref{fig_measures}}(b)  demonstrates the dependence of~$\Delta_{\omega}$  on the system size for three values of the stabilizing control parameter $K_s$. We keep the coupling radius $r=R/N=1/3$ fixed.  For small systems,  $\Delta_{\omega}$ increases with the system size, followed by saturation of its value for larger system size. Optimum control for $\varepsilon=0.2$ occurs at  $K_s=0.5$ in accordance with Fig.~5(a) of the main paper.


\subsection{Systems of inhomogeneous oscillators}
\numberwithin{equation}{subsection}
\setcounter{equation}{0}
\begin{figure*}[Ht!]
\includegraphics[height=\linewidth, angle=270]{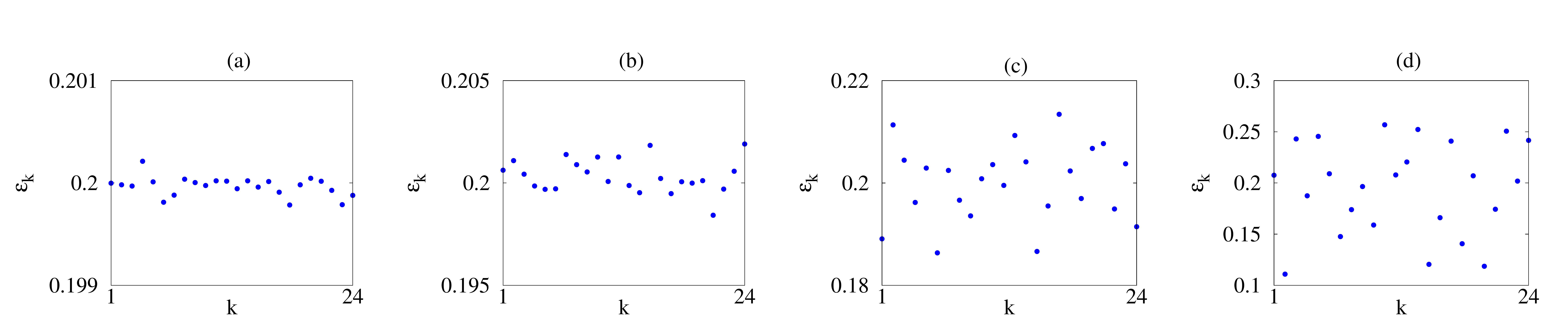}
\caption{(Color online) Exemplary realizations of $\varepsilon_k$ for $\varepsilon_{mean}=0.2$ and standard deviations (a)~$\delta_{\varepsilon}=0.0001$; (b)~$\delta_{\varepsilon}=0.001$; (c)~$\delta_{\varepsilon}=0.01$, (d)~$\delta_{\varepsilon}=0.05$.}
\label{fig_eps_distr}
\end{figure*}
\begin{figure*}[Ht!]
\includegraphics[height=\linewidth, angle=270]{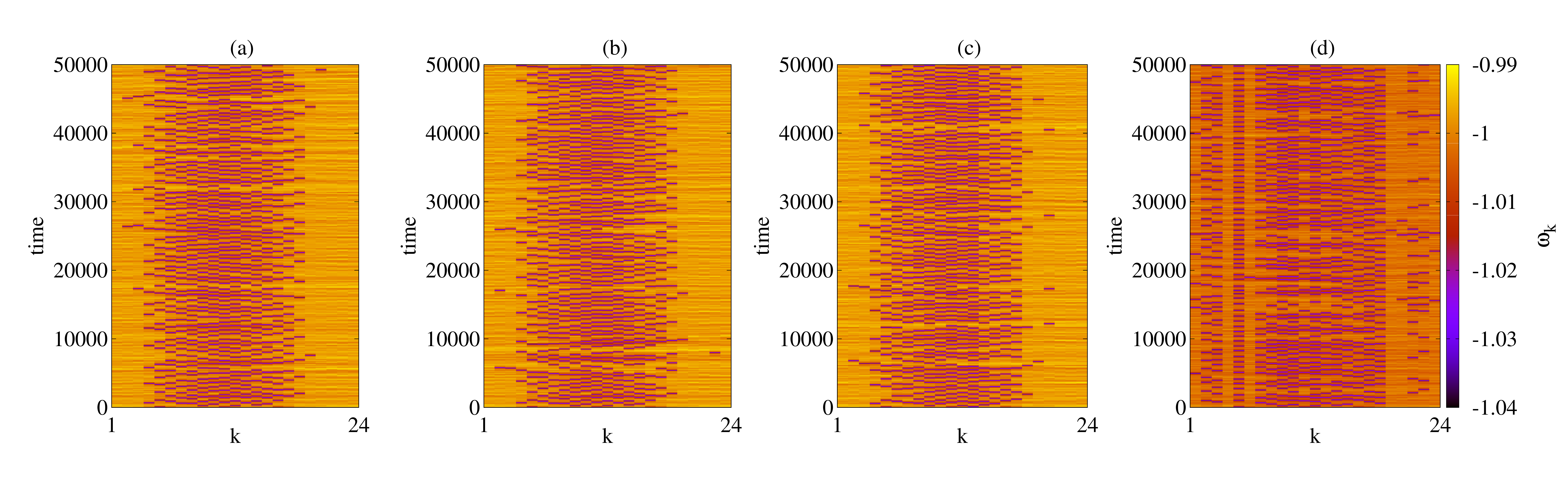}
\caption{(Color online) Mean phase velocities for a system of $N=24$ oscillators,
and $R=8$, $a=0.02$, $K_s=0.5$, $K_a=2$. Parameters $\varepsilon_k$   are drawn from a Gaussian distribution with $\varepsilon_{mean}=0.2$ as shown in the corresponding panels of Fig.~\ref{fig_eps_distr}(a)-(d).}
\label{fig_inhomogenVDP}
\end{figure*}
To prove the robustness of our control scheme, we consider a system of inhomogeneous Van der Pol oscillators: 
\begin{eqnarray}
\ddot{x}_k &=& (\eps_k - x_k^2)\dot{x}_k - x_k \nonumber\\[-1mm]
&+& \dfrac{1}{R} \sum\limits_{j=1}^R \left[ a_{-} (x_{k-j}-x_k) + b_{-}(\dot{x}_{k-j} - \dot{x}_k) \right] \nonumber\\[-2mm]
&+& \dfrac{1}{R} \sum\limits_{j=1}^R \left[ a_{+} (x_{k+j}-x_k) + b_{+}(\dot{x}_{k+j} - \dot{x}_k)  \vphantom{\sum} \right]
\label{Eq:VdP_inhomogen}
\end{eqnarray}
The individual Van der Pol oscillators have nonidentical nonlinearity parameters $\varepsilon_k$ and hence different frequencies, where  $\varepsilon_k$ is chosen randomly from a normal (Gaussian) distribution with mean value $\varepsilon_{mean}$ and standard deviation $\delta_{\varepsilon}$.
We fix $\varepsilon_{mean}=0.2$, and vary $\delta_{\varepsilon}$ to increase the inhomogeneity.

Fig.~\ref{fig_eps_distr} demonstrates exemplary realizations of $\varepsilon_k$ with increasing width of the Gaussian distribution.
Fig.~\ref{fig_inhomogenVDP} shows the mean phase velocities for the system~(\ref{Eq:VdP_inhomogen}) of $N=24$ Van der Pol oscillators, where $\varepsilon_k$ is taken from Fig.~\ref{fig_eps_distr}. For small inhomogeneity, the controlled chimera state is robust. With increasing inhomogeneity, chaotic elements start to appear in the coherent domain, leading eventually to its destruction for large inhomogeneity as shown in Fig.~\ref{fig_inhomogenVDP}(d).

\onecolumngrid
~~~
\twocolumngrid

\subsection{Control of chimeras in small networks of FitzHugh-Nagumo oscillators}
\numberwithin{equation}{subsection}
\setcounter{equation}{0}
As another example illustrating our control technique we consider
a system of~$N$ coupled identical FitzHugh-Nagumo oscillators
\begin{eqnarray}
\dot{X}_k = F_{\varepsilon,a}(X_k)
& + & \dfrac{1}{R} \sum\limits_{j=1}^R B_- ( X_{k-j} - X_k ) \nonumber\\[1mm]
& + & \dfrac{1}{R} \sum\limits_{j=1}^R B_+ ( X_{k+j} - X_k ),
\label{System:FHN}
\end{eqnarray}
where $X_k = (u_k,v_k)^{\mathrm{T}}\in\mathbb{R}^2$ is the state vector of the $k$-th oscillator and
\begin{equation}
F_{\varepsilon,a}(X_k) = \left(
\begin{array}{c}
( u_k - \frac{1}{3}u_k^3 - v_k ) / \varepsilon \\[2mm]
u_k + a
\end{array}
\right)
\end{equation}
is given by the nonlinear local dynamics of the FitzHugh-Nagumo model
with time-scale parameter $\varepsilon > 0$ and threshold parameter $a\in(-1,1)$ in the oscillatory regime, i.e.,
 each uncoupled oscillator exhibits a stable periodic orbit on a limit cycle.
Similar to Eq.~(1) in the main paper, we assume that each oscillator
is coupled with $R$~left and $R$~right nearest neighbors
such that the matrices~$B_-, B_+\in\mathbb{R}^{2\times 2}$
describe the local topology of coupling to the left and right, respectively.

The case of symmetric coupling
$$
B_- = B_+ = b S(\psi),
$$
where $b\in\mathbb{R}_+$ and
\begin{equation}
S(\psi) =
\left(
\begin{array}{ccc}
\cos \psi  & & \sin \psi \\
-\sin \psi  & & \cos \psi
\end{array}
\right)
\label{Matrx:B}
\end{equation}
is a rotational matrix with coupling phase $\psi$, has been considered in~\cite{S_OME13}.
There chimera states have been found for $\psi\lessapprox \pi/2$, 
and it has been shown that in the limit of small coupling strength $b << 1$ the phase dynamics of system~(\ref{System:FHN}) is approximately described by 
$$
\dot{\theta}_k = - \sum\limits_{j=-R}^R \sin(\theta_k - \theta_{k+j} + \alpha)
$$
where $\alpha\approx\psi$.

\begin{figure}[Ht!]
\includegraphics[height=1.05\linewidth, angle=270]{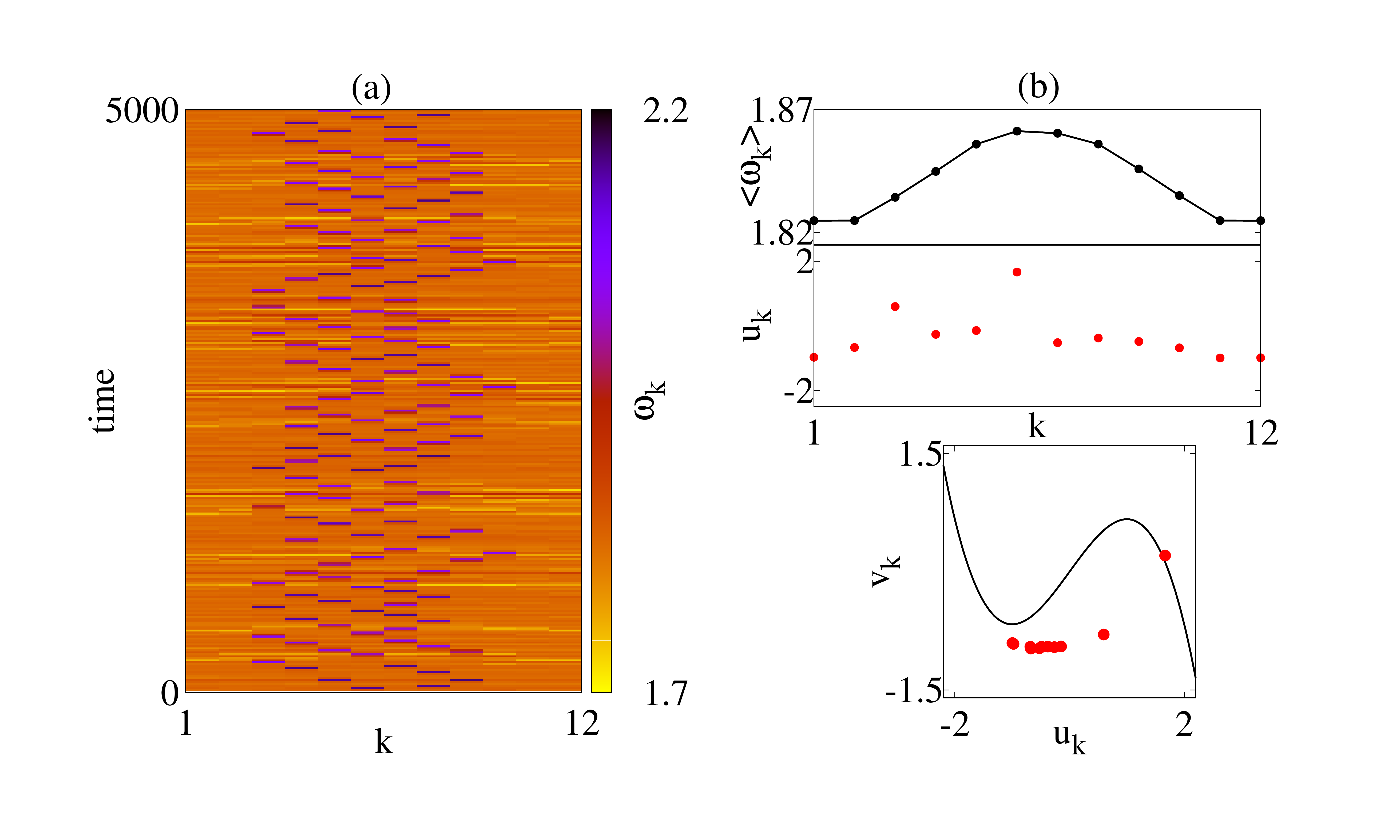}
\caption{(Color online) Controlled chimera state in the system (\ref{System:FHN})-(\ref{Psi_minus_plus}) of $N=12$ coupled FitzHugh-Nagumo oscillators.
Parameters: $R=4$, $\varepsilon = 0.15$, $a = 0.5$, $b= 0.15$, $K_{\mathrm{s}} = 2.0$, $K_{\mathrm{a}} = 1.0$.}
\label{fig_FHN}
\end{figure}

This suggests the following control scheme for system~(\ref{System:FHN}).
\begin{equation}
B_- = b S(\psi_-)\quad\mbox{and}\quad B_+ = b S(\psi_+)
\label{B_minus_plus}
\end{equation}
where
\begin{equation}
\psi_\pm = \dfrac{\pi}{2} - K_{\mathrm{s}} \left(1 - \dfrac{|Z_1 + Z_2|}{2} \right) \mp K_{\mathrm{a}}(|Z_1| - |Z_2|),
\label{Psi_minus_plus}
\end{equation}
$$
Z_1(t) = \frac{1}{[N/2]} \sum\limits_{k=1}^{[N/2]} e^{i \phi_k(t)}
$$
$$
Z_2(t) = \frac{1}{[N/2]} \sum\limits_{k=1}^{[N/2]} e^{i \phi_{N-k+1}(t)},
$$
and $\phi_k(t)$ is the geometric phase of the $k$-th oscillator computed from
$$
e^{i \phi_k (t)} = \left( u_k^2(t) + v_k^2(t) \right)^{-1/2} \left( u_k(t) + i v_k(t)\right).
$$
Now, for an appropriate choice of control gains $K_{\mathrm{s}}$ and $K_{\mathrm{a}}$ in~(\ref{Psi_minus_plus}) we can stabilize chimera states in the system (\ref{System:FHN})-(\ref{Psi_minus_plus}) with a small number of oscillators. Figure~\ref{fig_FHN} shows an example of a stabilized chimera state in a network of $N=12$ FitzHugh-Nagumo oscillators. 

\end{document}